\documentclass[hdvipdfm]{ptephy_v1}%%%%%% to generate preprint number with ptep logo

\preprintnumber{XXXX-XXXX} %%% %%% Insert preprint number here
\usepackage{bm}
\begin{document}

\title{Excitation energy spectra of the $\Lambda_{c}$ and $\Lambda_{b}$ baryons 
in a finite-size diquark model}

%%%% To generate auto affiliation numbers please use \author{}\affil{} command

\author{Kento Kumakawa}
\author{Daisuke Jido}
\affil{Department of Physics,
Tokyo Metropolitan University,  1-1 Minami-Osakwa, 
Hachioji, Tokyo, 192-0397, Japan \email{jido@tmu.ac.jp}}

%\author{Insert first author name here}
%\affil{Insert first author address here \email{xxxx@xxxx.ac.jp}}
%
%\author{Insert second author name here}
%\affil{Insert second author address here}
%
%\author{Insert third author name here}
%\author[3]{Insert fourth author name here} %%% Use optional bracket [3] to change the respective address
%\affil{Insert third author address here}
%
%\author{Insert last author name here\thanks{These authors contributed equally to this work}}
%\affil{Insert last author address here}

%%% To include the collaborator name... Please use the command "\collaborator"
%%% For example: \collaborator{ATLAS Collaboration}

\begin{abstract}%
The excitation energies of the $\Lambda_{c}$ and $\Lambda_{b}$ baryons are investigated
in a finite-size diquark potential model, in which the heavy baryons are treated as bound 
states of a charm quark and a scalar-isoscalar diquark. The diquark is considered as a 
sizable object. The quark-diquark interaction is calculated as a sum of the quark-quark 
interaction which is assumed to be half of the quark-antiquark
interaction for the color singlet.  The potential parameters in the quark-antiquark interaction
are fixed so as to reproduce the charmonium spectrum. 
We find the diquark size to be 1.1 fm for the diquark mass 0.5 GeV/c$^{2}$ to reproduce
the $1p$ excitation energy of $\Lambda_{c}$. In this model, the $\Lambda_{c}$ and $\Lambda_{b}$
excitation spectra are reproduced well, while this model does not explain $\Lambda_{c}(2765)$,
whose isospin nor spin-parity are unknown yet. Thus, the detailed properties of $\Lambda_{c}(2765)$
is very important to the presence of the diquark in heavy baryons as a effective constituent. 
We also discuss the $\Xi_{c}$ spectrum with the scalar strange diquark. 

\end{abstract}

\subjectindex{D32, B69}

\maketitle

\section{Introduction}

Understanding hadron spectroscopy in terms of quark-gluon dynamics is one of the challenging issues 
in hadron physics. Particularly, finding effective elements of hadron 
is important for intuitive understanding the hadron structure and excitation modes. Constituent quark
is one of the successful ideas to explain the hadron structure. For instance, the anomalous magnetic moments 
of nucleon and its flavor partners can be calculated as the matrix element of the quark magnetic 
moment operator in terms of the baryonic state constructed by the spin-flavor configuration of quarks,
being reproduced well if one regards the constituents as a point like Dirac fermion with 
a mass being one-third of the nucleon. It is interesting that, although the dynamics inside the hadron is very complicated,
such a simple picture works. This may be because the constituent quark can be a effective 
degrees of freedom for the hadron structure as a quasi particle formed as a consequence of complicated 
field theoretical dynamics of quarks and gluons.
The low-lying excitation spectra of heavy quarkonia are also explained by the excitation of the constituent
quarks in a simple confinement potential, so-called linear-plus-Coulomb type potential.  
The diquark, which is a pair of two quarks, is a longstanding strong candidate of a constituent of a hadron~\cite{Ida:1966ev,Lichtenberg:1967zz,Anselmino:1992vg}.
The diquark is regarded as a particle-like object in this study.
Having color charge, the diquark cannot exist alone in vacuum.  
So far, the existence of diquark correlation inside hadrons has been pointed out by 
phenomenological findings in baryon spectroscopy, weak non-leptonic decays, parton distribution functions,
and fragmentation functions.  Particularly one expects the existence of the scalar diquark 
having flavor, spin, and color antisymmetric configuration, the so-called good diquark, 
thanks to the most attractive correlation in perturbative QCD and instanton-induced interaction~\cite{Jaffe:2004ph}.
The strong correlation in the scalar diquark has been observed in 
lattice calculations~\cite{Hess98,Babich,Orginos,Alexandrou}.

In order to investigate the presence of the strong diquark correlation in hadron phenomenologically,
we focus on the excited spectra of heavy baryons. So far, many studies have been done for 
baryon spectra in terms of quark-diquark models. 
The mass spectra of the $\ell = 1$ and $\ell=2$ excited states of non-strange baryons 
were studied based on a diquark-quark model~\cite{Goldstein:1979wba} by using the SU(6)$\otimes$O(3)
classification~\cite{Lichtenberg:1981pp} and focusing on the fine structure by spin-orbit interactions.
In a relativistic formulation~\cite{Lichtenberg:1982jp}, the ground states of spin 3/2 baryons were investigated.
The radial and orbital excitations of baryons were calculated 
in a diquark-quark model with a confinement potential reproducing meson spectra~\cite{Liu:1983us}, 
and there detailed analyses for light flavor baryon spectra were given. 
The mass spectra of the excited heavy baryons with one heavy quark were studied 
in a relativistic quark-diquark model in which diquark bound states were solved relativistically 
and inner structure of diquark was taken into account~\cite{Ebert:2005xj,Ebert:2007nw}.
A finite size of the diquark has been suggested in Refs.~\cite{Alexandrou,Imai:2014yxa}, and 
there the size can be as larger as 1 fm. 
The new observed $\Omega_{c}$ states were investigated in a strange diquark model~\cite{Wang:2017vnc}.
In Ref.~\cite{Kim:2011ut}, the ground state masses of $\Lambda$, $\Lambda_{c}$
and $\Lambda_{b}$ were calculated in a diquark QCD sum rule, in which 
the scalar $ud$ diquark is explicitly considered as a fundamental field in operator product expansion.
The QCD sum rule successfully reproduced the observed $\Lambda$'s masses with 
a ``constituent" diquark mass 0.4 GeV, having satisfyed the standard criteria for the QCD sum rule to work.

Baryons composed of one heavy quark and two light quarks are good systems to study the diquark correlation.
As reported in Refs.~\cite{Jaffe76-1}, the quark-antiquark configurations dominate in meson wavefunctions,
and the diquark components are rather suppressed, because the diquark correlation is weaker than the 
quark-antiquark correlation. Thus, once the 
antiquark is present close to the diquark, the diquark may be easily broken up and form a quark-antiquark pair. 
In light baryons, the diquark correlation may be important for their structure, but rearrangement of the diquark
with the rest quark can be also important due to the symmetry among light quarks. 

In our previous work~\cite{Jido:2016yuv}, we have investigated excited spectra of the $\Lambda_{c}$ and
$\Lambda_{b}$ baryons in a quark-diquark model. In this model, the $ud$ diquark is treated as a point-like 
particle with isospin 0 and spin 0, and the heavy quark and the diquark are bound in the linear-plus-Coulomb
type potential. It has been found that, if one uses the potential parameters which reproduce the charmonium 
spectrum, one obtains $1p$ excitation energy of $\Lambda_{c}$ much more than the observation, and
in order to reproduce the observed excitation energy, one has to reduce the string tension by half
even though the antiquark and diquark have the same color charge. It has been also reported 
that the size effect of the diquark could solve this problem. 

In this paper, we consider the size of the diquark and calculate the excitation energy 
of $\Lambda_{c}$ and $\Lambda_{b}$ by treating the diquark as a rigid rotor.
The quark-diquark interaction is calculated as convolution of the quark-quark interaction. 
We will see that the finite size effect reduces the quark-diquark interaction at a short 
distance. This makes the excitation energies smaller for higher partial waves. 
We will find that the diquark size $\rho \simeq 1.1$ fm reproduces the observed 
excitation energy of the $1p$ $\Lambda_{c}$ state, and with this size the $\Lambda_{c}$ and 
$\Lambda_{b}$ spectra are well reproduced. We also discuss the mass and size of the
strange diquark from the $\Xi_{c}$ mass spectrum.

\section{Formulation}
We take a diquark-quark model for the heavy baryon which is composed of one heavy quark
($c$ or $b$ quark) and two light quarks. In the diquark-quark model, assuming that 
two light quarks forms a scalar diquark with antisymmetric flavor configuration, which is so-called good diquark, 
and we calculate the spectrum of the heavy baryon as a two-body problem with a heavy quark and a diquark.  
In the present work, we consider the diquark to be a sizable object not a point particle, 
and treat the diquark as a rigid rotor with the moment of inertia $I=\rho^{2} m_{d}/4$ where $\rho$ 
is the size of the diquark (the distance between two light quarks) and $m_{d}$ is the diquark mass. 
The distance between the light quarks, $\rho$, is not a dynamical variable any more.  

In the center of mass system, the coordinate representation of the Hamiltonian operator 
for the heavy quark and diquark system is written by
\begin{equation}\label{eq:H}
  H = m_{h} + m_{d} %+ \frac{p_{\lambda}^{2}}{2\mu} 
  - \frac{\hbar^{2}}{2\mu}\frac{1}{r} \frac{d^{2}}{dr^{2}}r + \frac{ L_{\lambda}^{2}}{2\mu r^{2}}
  + V(\vec r, \hat \rho) + \frac{ L_{\rho}^{2}}{2I}
\end{equation}
where $m_{h}$ and $m_{d}$ are the masses of the heavy quark and the diquark, respectively, 
$\mu$ is the reduced mass of the diquark and the heavy quark,
$r$ and $ L_{\lambda}$ are the radius and the angular momentum of 
the relative coordinate, 
$\hat L_{\rho}$ is the angular momentum of the diquark and 
$V(\vec r, \hat \rho)$ represents the interaction between the diquark 
and the heavy quark which depends on the relative coordinate $\vec r$ and
the orientation of the diquark $\hat \rho$. 

In the present work, 
the interaction is given by the sum of the interactions between 
the heavy quark and the light quarks inside the diquark as
\begin{equation}
  V(\vec r, \hat \rho) = V_{qq}(\vec r - \vec \rho/2) + V_{qq}(\vec r + \vec \rho /2) \label{dqpot}
\end{equation}
with $\vec \rho = \rho \hat \rho$ and the quark-quark interaction $V_{qq}$ for the $\bar {\bm 3}$
color configuration.
We presume that the strength of the $\bar {\bm 3}$ interquark interaction $V_{qq}$ be 
a half of the color electric quark-antiquark interaction $V_{\bar qq}$ for the color singlet configuration, 
$V_{qq} = \frac12 V_{\bar qq}$, 
which is the case of the one-gluon exchange calculation. 
The quark and antiquark interaction is assumed to be spherical and given by
a Coulomb plus linear from \cite{Eichten:1974af,Mukherjee:1993hb}
\begin{equation}
  V_{\bar qq}(r) = - \frac{4}{3} \frac{\alpha}{r} \hbar c + k r + V_{0} \label{pot}
\end{equation}
with three parameters, $\alpha$, $k$ and $V_{0}$. The parameter $V_{0}$ adjusts the 
absolute value of the mass spectrum and is irrelevant in this work, because we are 
interested in the excitation energies measured from the ground state. 
A choice of the parameters, $\alpha = 0.4$ 
and $k=0.9$~GeV~fm$^{-1}$, works well to produce the charmonium and bottonium spectra
\cite{Quigg:1979vr,Jido:2016yuv}. It is also reported in Ref.~\cite{Jido:2016yuv} that
the calculation of the first excitation energies of the $D$, $D_{s}$, $B$ and $B_{s}$ mesons 
with these values of the parameters is consistent with the experimental observation. 
According to these phenomenological success to reproduce the meson spectra, we use the following values 
\begin{equation}
  \alpha = 0.4, \qquad 
  k= 0.9\ {\rm GeV/fm}
\end{equation}
for the potential parameters. 
In this study, because we are interested in the global structure of the baryon spectrum in the 
quark-diquark model, we do not take into account fine structure interactions, such as spin-orbit interaction,
for the present. 
Because we do not consider spin-dependent forces, the angular momentum of the system, $L$, is a
good quantum number to label the states. 
(For the spinless diquark, the total angular momentum of the system $J$ is given by $J=L\pm \frac{1}{2}$, 
and these states are to degenerate due to lack of the spin-orbit interaction.)

Now let us decompose the quantum state of the quark-diquark system $|\Psi \rangle$ in terms of 
the angular momentum $L$:
\begin{equation}
   | \Psi \rangle = |S \rangle + |P \rangle + | D \rangle + \cdots.
\end{equation}
Each angular momentum state $|L\rangle$ are written by the combination of the 
angular momentum states of the relative coordinate and the diquark, which we write 
as  $|\ell_{\rho}, \ell_{\lambda} \rangle_{L} $ with 
the diquark angular momentum $\ell_{\rho}$ and the relative angular momentum $\ell_{\lambda}$.
Because we consider the scalar diquark with asymmetry flavor configuration, symmetry 
allows the diquark angular momentum $\ell_{\rho}$ to be only an even number. Thus, the low 
angular momentum states are given as
\begin{eqnarray}
   |S \rangle &=& |0,0\rangle_{S} + | 2,2 \rangle_{S} + \cdots, \\
   |P \rangle &=& |0,1\rangle_{P} + | 2,1 \rangle_{P} + | 2,2 \rangle_{P} + \cdots, \\
   |D \rangle &=& |0,2\rangle_{D} + | 2,0 \rangle_{D} + | 2,1 \rangle_{D} + |2,2\rangle_{D} + \cdots.
\end{eqnarray}
The construction of the states is explained in Appendix~\ref{Astate}.

We solve the eigen equation $\hat H | L \rangle = M | L \rangle$ for each angular momentum state 
to obtain the baryon mass $M$ as a bound state of the heavy quark and diquark. 
For this purpose we calculate matrix element 
$_{L}\langle \ell_{\rho}^{\prime}, \ell_{\lambda}^{\prime}  | \hat H |  \ell_{\rho}, \ell_{\lambda}\rangle_{L}$ 
for each partial wave $L$ and diagonalize the Hamiltonian. As we shall see later, 
it will turn out that the off-diagonal elements are negligibly small and the diagonalization plays a minor role. 
Because state $ |  \ell_{\rho}, \ell_{\lambda}\rangle_{L}$ has definite angular momenta of the 
diquark and the relative motion, it is an eigenstate of $\hat L^{2}_{\rho}$ and $\hat L^{2}_{\lambda}$:
\begin{equation}
   \hat L^{2}_{\rho}  |  \ell_{\rho}, \ell_{\lambda}\rangle_{L} 
   = \ell_{\rho}(\ell_{\rho}+1) \hbar^{2} |  \ell_{\rho}, \ell_{\lambda}\rangle_{L}, \qquad
   \hat L^{2}_{\lambda}  |  \ell_{\rho}, \ell_{\lambda}\rangle_{L} 
   = \ell_{\lambda}(\ell_{\lambda}+1) \hbar^{2} |  \ell_{\rho}, \ell_{\lambda}\rangle_{L}
\end{equation}
Now writing the radial wavefunction as $R(r)$ and using orthogonality of the angular moment eigenstates, 
we obtain the matrix elements as
\begin{eqnarray}
\lefteqn{
   _{L}\langle \ell_{\rho}, \ell_{\lambda}  | \hat H |  \ell_{\rho}, \ell_{\lambda}\rangle_{L}
   = m_{h} + m_{d} } && \nonumber \\
  && + 
  \int r^{2} dr R^{*}(r) \left[
    - \frac{\hbar^{2}}{2\mu} \frac{1}{r} \frac{d^{2}}{dr^{2}} r 
    + \frac{\ell_{\lambda}(\ell_{\lambda}+1)\hbar^{2}}{2\mu r^{2}}
    + V_{\rm eff}(r) + \frac{\ell_{\rho}(\ell_{\rho}+1)\hbar^{2}}{2I}
   \right] R(r) \label{MEdiag}
\end{eqnarray}
for the diagonal elements %in which we have omitted the quark and diquark masses 
and
\begin{equation}
   _{L}\langle \ell_{\rho}^{\prime}, \ell_{\lambda}^{\prime}  | \hat H |  \ell_{\rho}, \ell_{\lambda}\rangle_{L}
   =   \int r^{2} dr R^{*}(r) V_{\rm eff}(r)  R(r)  \label{MEoff}
\end{equation}
for the off-diagonal elements. Here we have introduced the effective potential for the radial motion $V_{\rm eff}(r)$
which is calculated by 
integrating the quark-diquark interaction potential~(\ref{dqpot}) with respect to
the solid angle integrals of the relative coordinate $\Omega_{r}$ and 
the diquark orientation $\Omega_{\rho}$:
\begin{eqnarray}
  V_{\rm eff}(r) 
  &=&\ _{L}\langle \ell_{\rho}^{\prime}, \ell_{\lambda}^{\prime}  | V(\vec r, \hat \rho) |  \ell_{\rho}, \ell_{\lambda}\rangle_{L} \\
  &=& \int d\Omega_{r} d\Omega_{\rho} \left[
  Y_{\ell_{\rho}^{\prime}}^{*m_{\rho}^{\prime}}(\Omega_{\rho}) 
  Y_{\ell_{\lambda}^{\prime}}^{*m_{\lambda}^{\prime}}(\Omega_{r}) \right]_{L}
  V(\vec r, \hat \rho)
  \left[Y_{\ell_{\rho}}^{m_{\rho}}(\Omega_{\rho}) Y_{\ell_{\lambda}}^{m_{\lambda}}(\Omega_{r}) \right]_{L},
  \label{eq:Veff}
\end{eqnarray}
where $[\cdots]_{L}$ means that one should take the appropriate linear combination 
so as to make the combined angular momentum to be $L$.
The explicit expression for each matrix element is given in Appendix~\ref{Apot}. 
We numerically solve the radial Schr\"odinger
equation
\begin{equation}\label{Sheq}
   \left[- \frac{\hbar^{2}}{2\mu} \frac{d^{2}}{dr^{2}} 
   + \frac{\ell_{\lambda}(\ell_{\lambda}+1)\hbar^{2}}{2\mu r^{2}}
    + V_{\rm eff}(r) + \frac{\ell_{\rho}(\ell_{\rho}+1)\hbar^{2}}{2I}
    \right] \chi(r) = E \chi(r)
\end{equation}
with $R(r) = \chi(r) / r$, and obtain the diagonal matrix element as $m_{h}+m_{d} +E$. 
For the off-diagonal elements, we perform the integral (\ref{MEoff}) using the wavefunctions
which are obtained from the Schr\"odinger equation for the diagonal components. 

\section{Results}
In this section, we show the numerical results of our calculation. 
First we show the effective potential of the diquark-quark interaction for several 
diquark sizes. Here we see that the size of the diquark reduces the interaction 
strength in shorter distance.
Then, we present our results on the excitation energy spectrum of $\Lambda_{c}$ obtained by 
solving the radial Schr\"odinger equation (\ref{Sheq}) as a function of the diquark size $\rho$.
There we determine the diquark size so as to reproduce the excitation energy 
of $\Lambda_{c}$ with $\ell =1$ for the diquark mass $0.5$ GeV/c$^{2}$. 
Next we show the calculated energy spectra of $\Lambda_{c}$ and $\Lambda_{b}$
using the determined diquark size and compare with the experimental observation. 
Then, we investigate the diquark mass dependence of the diquark size 
which reproduces the $p$-wave excitation energy. Finally we calculate the $\Xi_{c}$ 
mass spectrum
with the strange diquark $qs$ and determine the strange diquark mass and size. 
In our calculation, the masses of the charm quark and the bottom quark are fixed
as $1.5$ GeV/c$^{2}$ and 4.0 GeV/c$^{2}$, and 
we use $\alpha=0.4$ and $k=0.9$ GeV/fm for the potential parameter which reproduce
the charmonium spectrum well.  We have checked that 
the qualitative feature of our results is insensitive to the fine-tuning of these parameters. 
%For the diquark mass, we take its mass as $0.5$ GeV/c$^{2}$,

%First, we show our results for the excitation energy spectra of ${\Lambda_{c}}$ gained by solving the radial Schr\"odinger equation (\ref{Sheq}).We find the diquark size $\rho$ by comparing the experimental value of ${\Lambda_{c}}$ with the calculated value of that.\\
%Second, We compare the experiment value of ${\Lambda_{c}}$ and $\Lambda_{b}$ and the calculate value of the excitation energy of that by fixing the diquark size ${\rho}$.\\
%Third, we show the relation between the diquark mass $m_{d}$ and the diquark size $\rho$ by changing $m_{d}$ instead of $\rho$.\\
%Last, we compare the experimental value of $\Xi_{c}$ and the calculated value of that using the diquark size $\rho$ obtained by $\Lambda_{c}$.First of all, we show at the $\Lambda_{c}$ picture whether the $\Xi_{c}$'s that can be explained.After that, we determine the strange diquark mass $m_{ds}$ and the diquark size $\rho$ from the comparison of the energies of the experiment value and the calculated value at 1s and 1p quantum state.

\subsection{Effective potential between diquark and heavy quark}
\begin{figure}[htbp]
\begin{center}
   \includegraphics[clip]{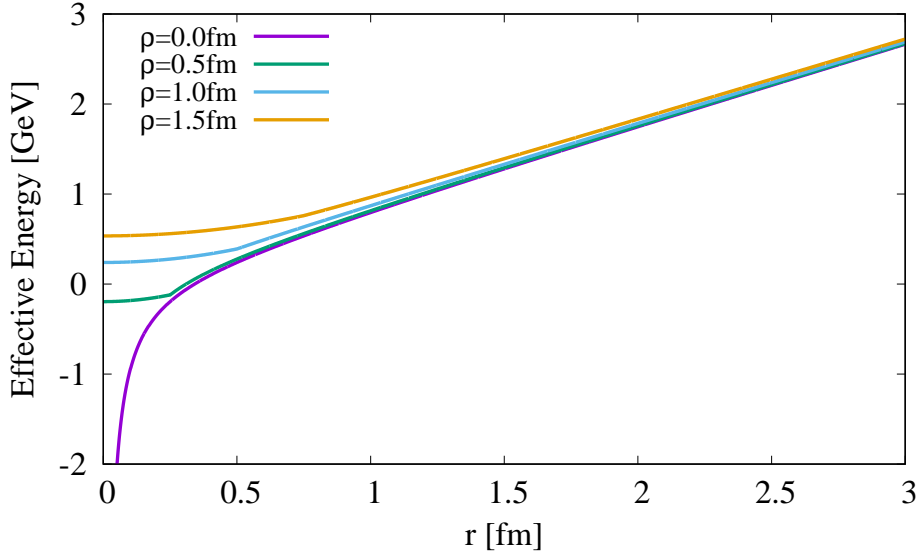}
\end{center}
\caption{Effective interaction potential between the heavy quark and the sizable diquark for 
lower energy states, $|0,0\rangle_{S}$, $|0,1\rangle_{P}$, $|0,2\rangle_{D}$ and $|2,0\rangle_{D}$,
which provide the equivalent effective potential. We plot the effective potentials for the diquark size,
$\rho =0$, 0.5, 1.0 and 1.5 fm.}  \label{fig:Veff}
\end{figure}

First of all, we discuss the interaction potential between the heavy quark and the sizable diquark.
The definition of the effective potential of the quark-diquark interaction is given in Eq.~(\ref{eq:Veff})
and the explicit calculation is done in Appendix~\ref{Apot}.
We are interested in the lower excitation spectrum. Here we show the effective potentials for 
the lower energy states.
In Fig.~\ref{fig:Veff}, we plot the effective potentials with several sizes of 
the diquark for the $|0,0\rangle_{S}$, $|0,1\rangle_{P}$, $|0,2\rangle_{D}$ and $|2,0\rangle_{D}$ states,
which are equivalent as shown in Eqs.~(\ref{eq:CS1}) to (\ref{eq:CD2}) and Eqs.~(\ref{eq:LS1}) to (\ref{eq:LD2})
of Appendix~\ref{Apot}. As seen in Fig.~\ref{fig:Veff}, at shorter distance $r < \rho/2$, 
the potential strength get reduced as a finite size effect. This will make the excitation energy smaller, because
the reduction of the attraction at short distance pushes the wave function out and,
for higher partial waves, this makes the effect of the centrifugal repulsion suppressed. 
In the following sections, we explicitly calculate the excitation energies of the heavy baryons. 

\subsection{Excitation energy of {${\Lambda_{c}}$} and determination of the diquark size ${\rho}$}
Let us show our result of the excitation spectrum of $\Lambda_{c}$ as a 
function of the diquark size. We calculate the energies of 
the $1s$, $1p$ , $1d$ and $2s$ states by solving 
the Schr\"odinger equation given in Eq. (\ref{Sheq}). The diquark mass is 
fixed to $m_{d}=0.5$ GeV/c$^{2}$. For the $1s$ and $2s$ states, we take 
the lowest angular momentum configuration 
$|\ell_{\rho}, \ell_{\lambda}\rangle_{L} = |0,0\rangle_{S}$
%$\ell_{\rho}=\ell_{\lambda}=0$,
and the $2s$ state is obtained as the first radial excitation state with $\ell=0$. 
For the $1p$ state, we take $|0,1\rangle_{P}$, which is 
the lowest angular momentum configuration for $\ell=1$. 
%$\ell_{\rho}=0$ and $\ell_{\lambda} =1$.
We have confirmed that the states with higher angular momentum configurations
are well separated and give essentially no effects to the lowest states. 
For the $1d$ state,
we take two states $| 0,2\rangle_{D}$ and $|2,0\rangle_{D}$.  
%$(\ell_{\rho},\ell_{\lambda})=(0,2)$ and $(2,0)$.
We consider the mixing of these two states by calculating off-diagonal 
matrix element of the effective potential (\ref{MEoff}), and obtain the energy eigenstates 
by diagonalizing the Hamiltonian. We call lower state as $1d_{1}$ and higher as $1d_{2}$.

\begin{figure}[htbp]
\begin{center}
   \includegraphics[clip]{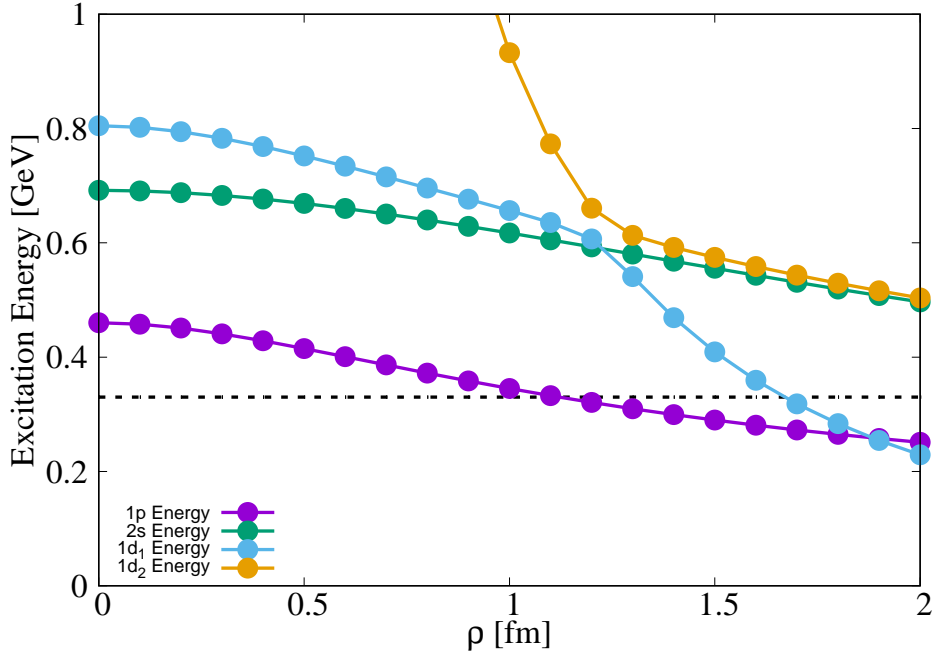}
\end{center}
\caption{Calculated $\Lambda_{c}$ excitation energies for the $1p$, $2s$ and $1d$
states as functions of the diquark size $\rho$. The energies are measured from 
the calculated $1s$ state energy. The dotted line stands for the spin weighted 
average of the observed excitation energies of $\Lambda_{c}(2595)$ and 
$\Lambda_{c}(2625)$, 0.330 GeV.   } \label{fig:1s1p}
\end{figure}

In Fig.~\ref{fig:1s1p} we show the calculated excitation energies of 
the $1p$, $2s$ and $1d$ states as a function of the diquark size $\rho$. 
These energies are measured from the calculated $1s$ energy.
The excitation energies decrease as the diquark size increases.
This is our expected result. For the point-like diquark, which is the case of
$\rho=0$ in the figure, the confinement potential reproducing the meson spectra
is so strong that the excitation energies are overestimated than the experiments. 
The size of the diquark reduces the strength of the interaction between the diquark
and heavy quark. 

With $\rho \simeq 1.2$ fm, level crossing takes place 
for the $|0,2\rangle_{D}$ and $|2,0\rangle_{D}$ states. The state $|2,0\rangle_{D}$ has
$\ell_{\rho}=2$, being a rotational state of the diquark. Thus, the excitation energy
is in inverse proportion to the diquark momentum of inertia and decreases as 
the diquark size increases. Because the states $|0,2\rangle_{D}$ and $|2,0\rangle_{D}$
have different angular momenta for the diquark-quark relative motion,
the wavefunctions are almost orthogonal. Thus, the mixing between these two 
states is negligibly small. 
 
%$(\ell_{\rho}, \ell_{\lambda}) = (0,2)$ and $(2,0)$ states with $\ell =2$.
%The $(\ell_{\rho}, \ell_{\lambda}) = (0,2)$ state

We determine the diquark size so as to reproduce the $1p$ excitation energy 
$\Delta E_{1p}$. In our calculation we do not consider the fine splittings 
caused by the spin-orbit interaction, since we are interested in the global 
feature of the diquark-quark interaction. Here we compare our results
with the experiments by taking spin weighted average of the observed 
masses for the $LS$ splitting partners, which removes the effect of 
the spin-orbit interaction in perturbation theory. For the $1p$ state, 
$\Lambda_{c}(2595)$ with $1/2^{-}$ and $\Lambda_{c}(2625)$ with $3/2^{-}$
are the $LS$ partners. The spin weighted average of the excitation energy is 
obtained by $\Delta E_{\rm ave} = \frac23 \Delta E_{3/2^{-}} + \frac13 \Delta E_{1/2^{-}} 
= 0.330$ GeV. 
From Fig.~\ref{fig:1s1p}, we find that the diquark size $\rho = 1.1$ fm reproduces 
the experimental value $0.330$ GeV. This is consistent with 
the finding in Refs.~\cite{Alexandrou,Imai:2014yxa}.

%%%%%%%%

\subsection{Excited energies of $\Lambda_{c}$ and $\Lambda_{b}$ with the determined
diquark size}

In the previous section, we have determined the diquark size as $\rho = 1.1$ fm for the diquark 
mass $m_{d}=0.5$ GeV/c$^{2}$ from the excitation energy 
of the $1p$ state. Here we discuss the excitation energies of the other states of $\Lambda_{c}$ and show
the excitation energy spectrum of $\Lambda_{b}$ with the determined diquark size. 

\begin{figure}[htbp]
\begin{center}
   \includegraphics[clip]{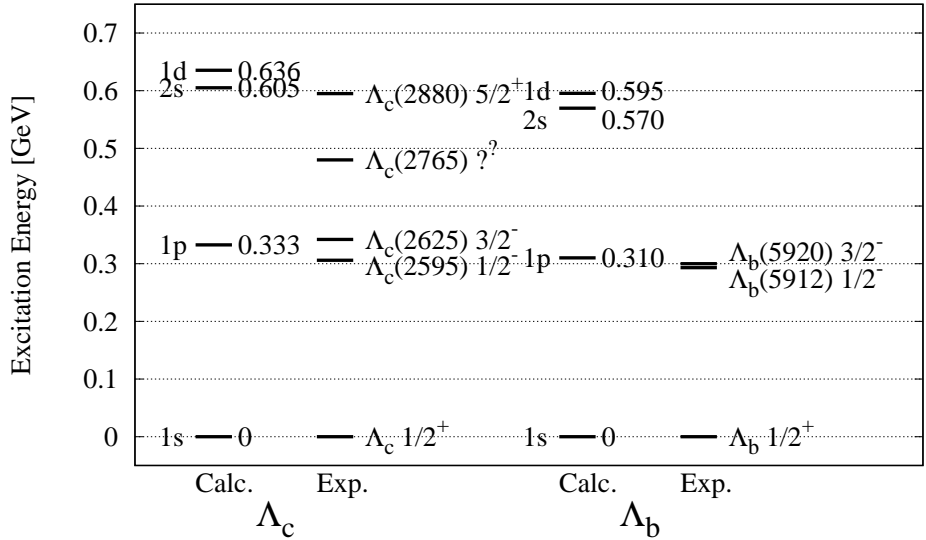}
\end{center}
\caption{Comparison of the calculated $\Lambda_{c}$ and ${\Lambda_{b}}$ excitation energies with 
the experimental data~\cite{PDG}. The calculation is done with the diquark mass $m_{d}=0.5$ GeV/c$^{2}$
and the diquark size $\rho = 1.1$ fm. 
The spin and parity of $\Lambda_{c}(2765)$ is known and its isospin is not fixed yet either. 
} 
\label{fig:LambdacLambdab}
\end{figure}

In Fig.~\ref{fig:LambdacLambdab}, we present the calculated result of the excitation energies of $\Lambda_{c}$
and $\Lambda_{b}$ with the diquark size $\rho = 1.1$ fm, 
%which reproduces the $1p$ excitation energy of $\Lambda_{c}$, 
and there we show also their possible corresponding observed states. 
As shown in the figure, the $2s$ states are obtained rather higher than usual quark model, in which 
$\Lambda_{c}(2765)$ is explained as the $2s$ state. This is because, in our model, 
the attraction between the diquark and heavy quark gets weaker when two particles 
are approaching, and thus this reduction of the attractive force is more effective for the $\ell=0$ states.
Our model fails to reproduce $\Lambda_{c}(2765)$.
The $\Lambda_{c}(2765)$ state is a one-star resonance in Particle Data~\cite{PDG}, and its spin and parity 
are unknown yet. Even its isospin is not fixed yet, so that it can be  $\Sigma_{c}$.
Therefore, the isospin of the $\Lambda_{c}(2765)$ state is the touchstone for the success of our model. 
If it would be $\Lambda_{c}$, the diquark picture could not be the case in $\Lambda_{c}$ excited states.
The $1d$ state is marginally reproduced but a bit higher than the observed $\Lambda(2880)$ state.

\subsection{The diquark mass dependence}
\begin{figure}[htbp]
\begin{center}
   \includegraphics[clip]{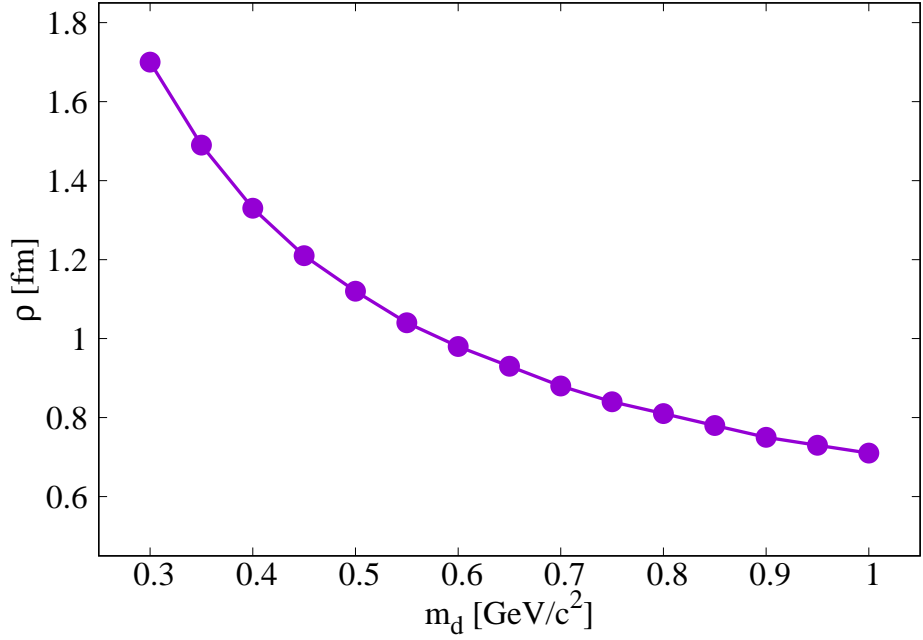}
\end{center}
\caption{Diquark mass dependence of the diquark size which reproduces 
the $1p$ excitation energy of $\Lambda_{c}$.} \label{fig:diquarkmass}
\end{figure}

Here we discuss the diquark mass dependence of the diquark size which reproduces the $1p$ excitation 
energy of $\Lambda_{c}$.  In Fig.~\ref{fig:diquarkmass}, we show the diquark size appropriate for 
the $1p$ excitation of $\Lambda_{c}$ as a function of the diquark mass. As seen in the figure, 
if one uses a lighter diquark, the size of the diquark should be larger 
to reproduce the $1p$ state of $\Lambda_{c}$. Because it is not likely that the diquark size is much 
larger than 1 fm, a possible diquark mass is larger than 0.5 GeV/c$^{2}$.

\begin{figure}[htbp]
\begin{center}
   \includegraphics[clip]{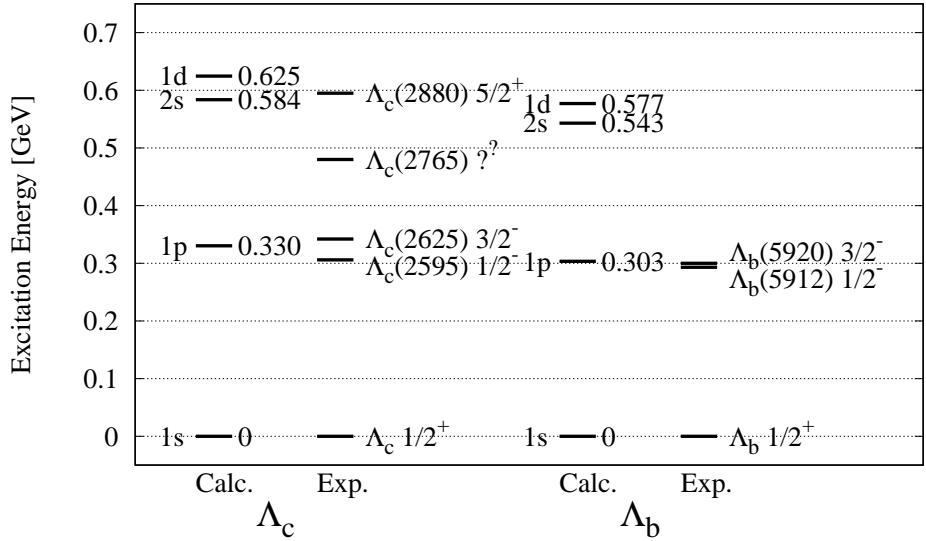}
\end{center}
\caption{Comparison of the calculated $\Lambda_{c}$ and ${\Lambda_{b}}$ excitation energies with 
the experimental data~\cite{PDG}. 
The calculation is done with the diquark mass $m_{d}=0.7$ GeV/c$^{2}$ and diquark size $\rho = 0.88$ fm. 
} 
\label{fig:LambdacLambdabmd07}
\end{figure}

In Fig.~\ref{fig:LambdacLambdabmd07}, we show the calculated $\Lambda_{c}$ and $\Lambda_{b}$
excitation energy spectra with the diquark mass $m_{d}=0.7$ GeV/c$^{2}$ and diquark size
$\rho=0.88$ fm which reproduce the $1p$ excitation energy of $\Lambda_{c}$. As seen in the figure,
The observed excitation energies are reproduced well with this parameter set and this 
is essentially same as the calculation with $m_{d}=0.5$ GeV/c$^{2}$, while one sees that
with the heavier diquark mass the excitation energies are a bit smaller. 

%if one uses a heavier  diquark, the size of the diquark should be more compact 
%to reproduce the $1p$ state of $\Lambda_{c}$. 

\subsection{${\Xi_{c}}$ energy spectrum }
In this section we discuss the energy spectrum of $\Xi_{c}$, which is composed of
one charm quark and one strange diquark. The strange diquark is formed by
one strange quark and one light (up or down) quark, having spin zero and antisymmetric 
flavor and color configurations. The values of the potential parameters appearing in Eq.~(\ref{pot}) 
are to be same with those used in the $\Lambda_{c}$ calculation. 

\begin{figure}[htbp]
\begin{center}
   \includegraphics[clip]{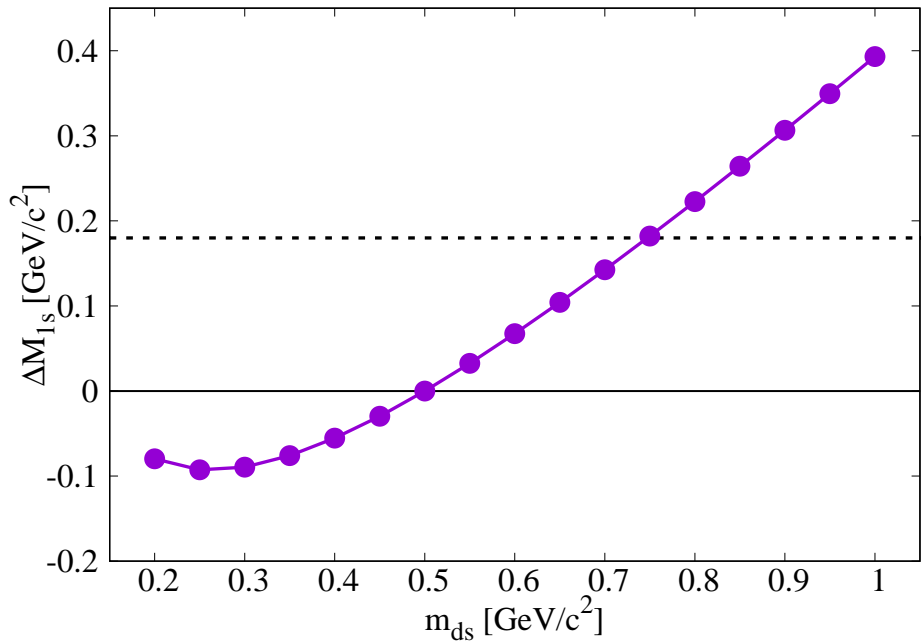}
\end{center}
\caption{Mass difference between the ground states of $\Xi_{c}$ and $\Lambda_{c}$ as
a function of the strange diquark mass. The size of the strange diquark mass is assumed to be same
as the $ud$ diquark size, which is $\rho =1.1$ fm. The mass of the $ud$ diquark is fixed as 0.5 GeV/c$^{2}$.
The horizontal dashed line stands for the observed value of the mass difference of $\Xi_{c}$ and $\Lambda_{c}$.
} \label{fig:mdiff}
\end{figure}

First of all, we determine the strange diquark mass from the mass difference between
the ground states of $\Lambda_{c}$ and $\Xi_{c}$. Here we assume the same
size for the strange diquark as the $ud$ diquark. 
The $ud$ diquark mass and diquark size are fixed as 0.5 GeV/c$^{2}$ and 1.1 fm, respectively. 
In Fig.~\ref{fig:mdiff}, we show the $\Xi_{c}$ - $\Lambda_{c}$ mass difference for the 
ground states as a function of the strange diquark mass, and the dashed line indicates
the observed value of the mass difference, 0.18 GeV/c$^{2}$. As shown in the figure, 
the ground state mass increases as the diquark mass increases. We find that 
with the strange diquark mass $m_{ds} = 0.75$ GeV/c$^{2}$ the observed mass of the 
lowest lying $\Xi_{c}$ state is reproduced. With this strange diquark mass we calculate
the $1p$ excitation energy and obtain 0.30 GeV. This is smaller than the experimental value 0.34~GeV,
which is obtained as the spin weighted average of the excitation energies of $\Xi_{c}(2790)$ and $\Xi_{c}(2815)$. 

%\newpage

\begin{figure}[htbp]
\begin{center}
   \includegraphics[clip]{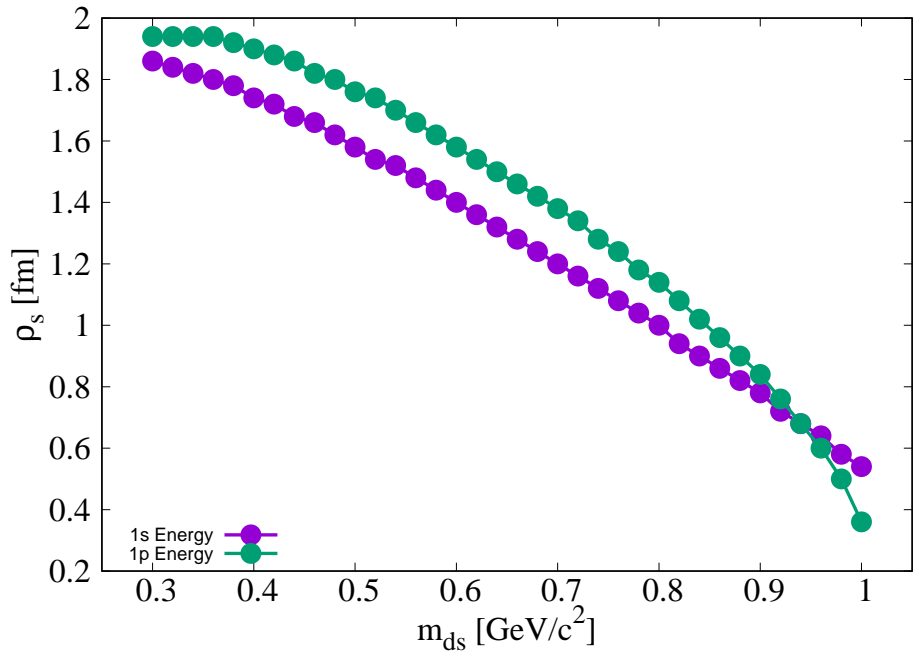}
\end{center}
\caption{The diquark mass and size reproducing the masses of the $1s$ and $1p$ states of $\Xi_{c}$. 
The plot shows that strange diquark mass $m_{ds}=0.94$ GeV/c$^{2}$ and the strange diquark size 
$d_{s}=0.68$ fm reproduce both the $1s$ and $1p$ masses of $\Xi_{c}$. }
\label{fig:xic}
\end{figure}

\begin{figure}[htbp]
\begin{center}
   \includegraphics[clip]{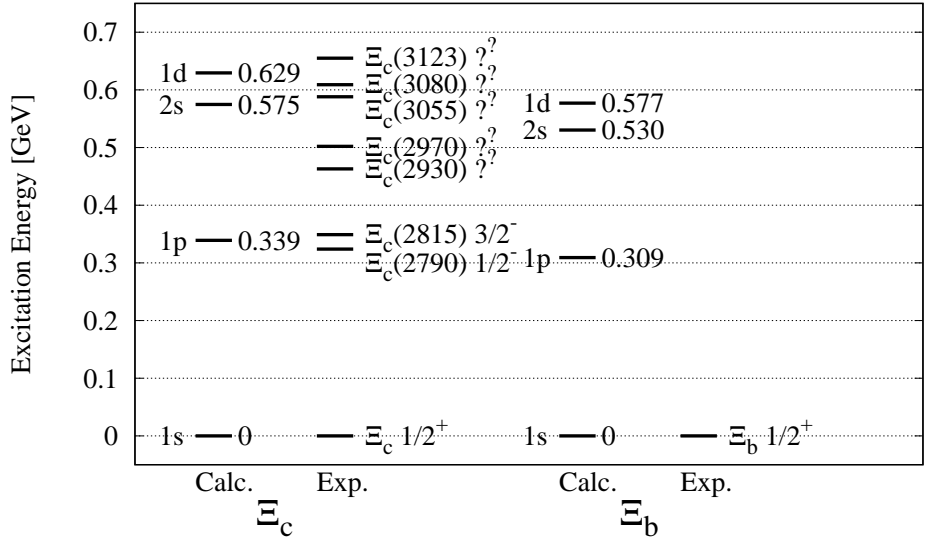}
\end{center}
\caption{Comparison of the calculated $\Xi_{c}$ excitation energies with 
the experimental data~\cite{PDG}. 
The calculation is done with the strange diquark mass $m_{ds}=0.94$ GeV/c$^{2}$ 
and strange diquark size $\rho_{s} = 0.68$ fm. We omit $\Xi^{\prime}$ and $\Xi(2645)$
in this figure, because the strange diquark may have spin 1 in these baryons. 
} 
\label{fig:Xic}
\end{figure}

Next let us determine both the strange diquark mass and size from the masses of the ground and 
first excited states of $\Xi_{c}$. We fix the parameter $V_{0}$ appearing in Eq.~(\ref{pot}) by
the lowest lying $\Lambda_{c}$ state with the diquark mass 0.5 GeV/c$^{2}$ and the diquark size
1.1 fm. For the $1p$ state of $\Xi_{c}$, the spin weighted average mass of 
$\Xi_{c}(2790)$ and $\Xi_{c}(2815)$ gives 2.809 GeV/c$^{2}$.
In Fig.~\ref{fig:xic}, we show the strange diquark sizes $\rho_{s}$ which reproduce the masses of 
the $1s$ and $1p$ states of $\Xi_{c}$ as functions of the strange diquark mass $m_{ds}$. 
From this figure, we find that $m_{ds} = 0.94$ GeV/c$^{2}$ and $\rho_{s}=0.68$ fm reproduce
both the masses of the $1s$ and $1p$ states. In Fig.~\ref{fig:Xic} we show the calculated 
$\Xi_{c}$ excitation energies with these diquark mass and size. Due to lack of the experimental 
information on the quantum numbers for the higher $\Xi_{c}$ states, it is not easy to make
further comparison.

\section{Conclusion}

We have investigated the excitation energy spectra of $\Lambda_{c}$ and $\Lambda_{b}$
in a finite-size diquark model, in which the heavy baryon is composed of one heavy quark 
and one diquark with a finite size. The diquark is treated as a rigid rotor of two light quarks
separated in the distance $\rho$. The interaction between the heavy quark and the diquark 
is calculated as sum of the heavy quark - light quark interactions for the $\bar {\bf 3}$
color configuration, which is assumed to be half of the quark - antiquark interaction
for the color singlet pair. The potential parameter is fixed so as to reproduce the charmonium 
spectrum with the quark - antiquark interaction. 

With the diquark size the interaction between the diquark and heavy quark is reduced in short
distance. This makes the excitation energies for the $\ell > 0$ states smaller, because
the weaker attraction pushes the wave function out and the effect of the centrifugal 
repulsion is suppressed. 
The present model reproduces the $1p$ excitation energy of $\Lambda_{c}$  
with the diquark mass 0.5 GeV/c$^{2}$ and the diquark size 1.1 fm. This diquark size is consistent with
the calculation in a Schwinger-Dyson formalism~\cite{Imai:2014yxa}.
The quark-diquark model with the diquark size 1.1 fm reproduces well the mass spectra
of $\Lambda_{c}$ and $\Lambda_{b}$. This model, however, does not provide 
$\Lambda_{c}(2765)$, which is usually assigned to the $2s$ state in quark models. 
Nevertheless, $\Lambda_{c}(2765)$ is such an uncertain state that its spin-parity is not 
known nor its isospin is not fixed yet. Thus, the detailed information on $\Lambda_{c}(2765)$
should be necessary for the diquark picture of the heavy baryon. 
We also discuss the energy spectrum of $\Xi_{c}$ with the strange diquark. 
The masses of the $1s$ and $1p$ states of $\Xi_{c}$ can be reproduced with 
the strange diquark mass $m_{ds} = 0.94$ GeV/c$^{2}$ and its size $\rho_{s}=0.68$ fm.

The calculation done in the present work is quite simple. Nevertheless, the quantitative feature 
should be reproduced by a simple insight. Thus, the presence of the diquark inside the baryons 
as a effective constituents should be also reproduced by simple models. If this is not the case,
we should be to give up diquark to explain the excitation mode in the $\Lambda_{c}$ baryon. 
In this sense the details information of $\Lambda_{c}(2765)$ is quite important. 

\section*{Acknowledgments}

The work of D.J.\ was partly supported by Grants-in-Aid for Scientific Research from JSPS (17K05449). 


%\bibliographystyle{ptephy}
%\bibliography{sample}
%
% once the .bbl file has been generated then place the text in your article.

\appendix
\section{Construction of states}  \label{Astate}
Each angular momentum state $|L\rangle$ is a linear combination of the direct products of 
the angular momentum states for the diquark and the relative motion. We write this product state
as $| \ell_{\rho}m_{\rho}\, \ell_{\lambda}m_{\lambda} \rangle$, where 
$\ell_{\rho}$ and $m_{\rho}$ ($\ell_{\lambda}$ and $m_{\lambda}$) are 
the orbital angular momentum and magnetic quantum numbers of 
the diquark (relative coordinate), respectively. The state with the angular momentum $L$ 
for the quark-diquark system combined by the angular momentum states of the diquark
and the relative motion with the quantum numbers $ (\ell_{\rho}, m_{\rho})$ and $(\ell_{\lambda}, m_{\lambda})$,
respectively, is written with the Clebsch-Gordan coefficient as
\begin{equation}
   | \ell_{\rho}, \ell_{\lambda} \rangle_{L} 
   = \sum_{m_{\rho}, m_{\lambda}} | \ell_{\rho}m_{\rho}\, \ell_{\lambda}m_{\lambda} \rangle
   \langle \ell_{\rho}m_{\rho}\, \ell_{\lambda}m_{\lambda} | L M \rangle
\end{equation}
The states are written explicitly as
\begin{eqnarray}
   |0,0 \rangle_{S} &=& |00\,00\rangle \\
   |2,2 \rangle_{S} &=& \frac{1}{\sqrt 5} \left(
    | 2 {\, -2}\, 2 {\, +2} \rangle -| 2 {\, -1}\, 2 {\, +1} \rangle + | 20\,20 \rangle - | 2 {\, +1}\, 2 {\, -1} \rangle
    + | 2 {\, +2}\, 2 {\, -2} \rangle\right)
\end{eqnarray}
for the $S$ states, 
\begin{eqnarray}
    |0,1\rangle_{P} &=& |00\,10 \rangle\\
    |2,1\rangle_{P} &=& \sqrt{\frac{3}{10}} |2{\,-1}\, 1{\,+1}\rangle - \sqrt{\frac25} |20\,10\rangle 
    + \sqrt{\frac{3}{10}} | 2{\,+1}\, 1{\,-1}\rangle \\
    |2,2 \rangle_{P} &=& \sqrt\frac25 | 2 {\, -2}\, 2 {\, +2} \rangle - \sqrt\frac1{10} | 2 {\, -1}\, 2 {\, +1} \rangle 
    + \sqrt\frac1{10} | 2 {\, +1}\, 2 {\, -1} \rangle - \sqrt\frac25 | 2 {\, +2}\, 2 {\, -2} \rangle
\end{eqnarray}
for the $P$ states, and
\begin{eqnarray}
    |0,2\rangle_{D} &=& |00\, 20\rangle \\
    |2,0\rangle_{D} &=& |20\, 00\rangle \\
    |2,1\rangle_{D} &=& \frac{1}{\sqrt 2} \left( |2{\, +1}\, 1{\, -1} \rangle - | 2{\, -1}\, 1{\, +1}\rangle \right)\\
    |2,2 \rangle_{D} &=& \sqrt\frac27 | 2 {\, -2}\, 2 {\, +2} \rangle + \sqrt\frac1{14} | 2 {\, -1}\, 2 {\, +1} \rangle 
    - \sqrt\frac27 | 20\,20 \rangle  \nonumber \\ && 
    + \sqrt\frac1{14} | 2 {\, +1}\, 2 {\, -1} \rangle + \sqrt\frac27 | 2 {\, +2}\, 2 {\, -2} \rangle
\end{eqnarray}
for the $D$ states.

\section{Effective potential for quark and diquark interaction}  \label{Apot}
The effective potential appearing in Eqs.~(\ref{MEdiag}) and (\ref{MEoff}) is 
a linear combination of the matrix element of the interaction potential $V(\vec r, \rho)$ 
for the angular momentum state $| \ell_{\rho}m_{\rho}\, \ell_{\lambda}m_{\lambda} \rangle$.
Here we derive the effective potentials for the Coulomb type potential 
$\displaystyle{v(\vec r_{1} - \vec r_{2}) = \frac{1}{|\vec r_{1} - \vec r_{2}|}}$
%$V(\vec r_{1} - \vec r_{2}) = \frac{1}{|\vec r_{1} - \vec r_{2}|}$
%$V(\vec r_{1} - \vec r_{2}) = 1/|\vec r_{1} - \vec r_{2}|$
and linear type potential $\displaystyle{v(\vec r_{1}- \vec r_{2}) = |\vec r_{1} - \vec r_{2}|}$. 

To obtain the effective potential, we perform the Legendre expansion for the interaction potential 
$v(\vec r_{1} - \vec r_{2})$ in terms of the relative angle $\alpha$ of $\vec r_{1}$ and $\vec r_{2}$.
\begin{equation}
   v(\vec r_{1}-\vec r_{2}) = \sum_{\ell=0}^{\infty} v_{\ell}(r_{1},r_{2}) P_{\ell}(\cos\alpha)
\end{equation}
The Legendre coefficients for the Coulomb and linear type potentials are known as 
\begin{eqnarray}
   v_{\ell}(r_{1},r_{2}) = \left\{
   \begin{array}{ll}
    \displaystyle{ \frac{1}{r_{>}} \left( \frac{r_{<}}{r_{>}} \right)^{\ell} } &
     {\rm for} \  \displaystyle{ v = \frac{1}{|\vec r_{1} - \vec r_{2}|} } \\
     \displaystyle{ r_{>} \left[ \frac{1}{2\ell + 3} \left(\frac{r_{<}}{r_{>}}\right)^{\ell+2}
      - \frac{1}{2\ell-1} \left( \frac{r_{<}}{r_{>}} \right)^{\ell}\right]}&
     {\rm for} \  \displaystyle{ v =|\vec r_{1} - \vec r_{2}| } 
   \end{array}
   \right.
\end{eqnarray}
where $r_{>}$ ($r_{<}$) denotes the greater (less) of $|\vec r_{1}|$ and $|\vec r_{2}|$.

Using the addition theorem for the Legendre polynomials, we write down the potential 
with the spherical harmonics for the solid angles of $\vec r_{1}$ and $\vec r_{2}$:
\begin{equation}
   v(\vec r_{1}- \vec r_{2}) = \sum_{\ell=0}^{\infty} v_{\ell}(r_{1},r_{2}) \frac{4\pi}{2\ell +1}
    \sum_{m=-\ell}^{\ell} Y_{\ell}^{m*}(\Omega_{1}) Y_{\ell}^{m}(\Omega_{2})
\end{equation}
The matrix element of $v(\vec r_{1}- \vec r_{2})$ of the state $| \ell_{1}m_{1}\, \ell_{2}m_{2} \rangle$
can be calculated as
\begin{eqnarray}
\lefteqn{
  \langle \ell_{1}^{\prime}m_{1}^{\prime}\, \ell_{2}^{\prime} m_{2}^{\prime} |
   v(\vec r_{1}- \vec r_{2}) | \ell_{1}m_{1}\, \ell_{2}m_{2}\rangle
} && \nonumber \\
&=&  \sum_{\ell=0}^{\infty} v_{\ell}(r_{1},r_{2}) \sum_{m=-\ell}^{\ell} 
    \langle \ell_{1}^{\prime} m_{1}^{\prime}|C_{\ell}^{m*}(\Omega_{1}) | \ell_{1} m_{1} \rangle
    \langle \ell_{2}^{\prime} m_{2}^{\prime} | C_{\ell}^{m}(\Omega_{2}) | \ell_{2} m_{2} \rangle,
\end{eqnarray}
where we have defined
\begin{equation}
    C_{\ell}^{m}(\Omega) = \sqrt \frac{4\pi}{2\ell+1} Y_{\ell}^{m}(\Omega).
\end{equation}
The matrix element of $C_{\ell}^{m}$ is written in terms of the $3j$ symbol as
\begin{eqnarray}
  %\lefteqn{
  \langle \ell^{\prime} m^{\prime}|  C_{L}^{M}| \ell m \rangle 
  %} \nonumber \\ 
  & =&  \sqrt \frac{4\pi}{2L+1}  \int d\Omega 
  Y^{m^{\prime}*}_{\ell^{\prime}}(\Omega) Y^{m}_{\ell}(\Omega) Y_{L}^{M}(\Omega) \\
    &=&  (-)^{-m^{\prime}}\sqrt{(2\ell^{\prime} +1)(2\ell+1)}
  \left(\begin{array}{ccc} \ell^{\prime}& \ell & L \\ 0 & 0 & 0\end{array}\right) 
  \left(\begin{array}{ccc} \ell^{\prime}& \ell & L \\ -m^{\prime} & m & M\end{array}\right) %\nonumber
\end{eqnarray}
The $3j$ symbol is defined by Clebsch-Gordan coefficient as
\begin{equation}
   \left(\begin{array}{ccc} j_{1}& j_{2} & J \\ m_{1} & m_{2} & -M\end{array}\right) 
   = \frac{(-)^{j_{1}-j_{2}+M}}{\sqrt{2J+1}} \left\langle j_{1} j_{2} m_{1} m_{2} | JM \right \rangle
\end{equation}
The matrix element has the following properties:
\begin{eqnarray}
 \langle \ell^{\prime} m^{\prime}|  C_{L}^{M*}| \ell m \rangle
 &=& (-)^{M} \langle \ell^{\prime} m^{\prime}|  C_{L}^{-M}| \ell m \rangle \\
 \langle \ell^{\prime} m^{\prime}|  C_{L}^{M}| \ell m \rangle
 &=& (-)^{m^{\prime} + m} \langle \ell m|  C_{L}^{M}| \ell^{\prime} m^{\prime} \rangle 
\end{eqnarray}

In the followings we list the effective potentials of the Coulomb type potential for each partial wave:
\begin{eqnarray}
  _{S}\langle 0,0 | \frac{1}{|\vec r_{1} - \vec r_{2}|} | 0,0 \rangle_{S}
  &=& 
  %\langle 00\,00 | \frac{1}{|\vec r_{1} - \vec r_{2}|} | 00\,00 \rangle = 
  \frac{1}{r_{>}}   \label{eq:CS1} \\
%\end{eqnarray}
%\begin{eqnarray}
  _{P}\langle 0,1 | \frac{1}{|\vec r_{1} - \vec r_{2}|} | 0,1 \rangle_{P}
  &=&  
   %\langle 00\,10 | \frac{1}{|\vec r_{1} - \vec r_{2}|} | 00\,10 \rangle= 
   \frac{1}{r_{>}} \\
%\end{eqnarray}
%\begin{eqnarray}
  _{D}\langle 0,2 | \frac{1}{|\vec r_{1} - \vec r_{2}|} | 0,2 \rangle_{D}
  &=&  \frac{1}{r_{>}} \\
  _{D}\langle 2,0 | \frac{1}{|\vec r_{1} - \vec r_{2}|} | 2,0 \rangle_{D}
  &=&  \frac{1}{r_{>}}  \label{eq:CD2} \\
  _{D}\langle 2,0 | \frac{1}{|\vec r_{1} - \vec r_{2}|} | 0,2 \rangle_{D}
  &=& \frac15 \frac{1}{r_{>}} \left(\frac{r_{<}}{r_{>}}\right)^{2}  
\end{eqnarray}

We list also the effective potentials of the linear type potential:
\begin{eqnarray}
  _{S}\langle 0,0 || \vec r_{1} - \vec r_{2} || 0,0 \rangle_{S} 
         &=& r_{>}\left[1+ \frac{1}{3}\left(\frac{r_{<}}{r_{>}}\right)^{2} \right] \label{eq:LS1} \\
  _{P}\langle 0,1 || \vec r_{1} - \vec r_{2} || 0,1 \rangle_{P}
         &=& r_{>} \left[ 1 + \frac{1}{3}\left(\frac{r_{<}}{r_{>}}\right)^{2}  \right] \\
  _{D}\langle 0,2 | {|\vec r_{1} - \vec r_{2}|} | 0,2 \rangle_{D}
  &=& {r_{>}} \left[1 + \frac{1}{3} \left(\frac{r_{<}}{r_{>}}\right)^{2}\right]  \\
  _{D}\langle 2,0 | {|\vec r_{1} - \vec r_{2}|} | 2,0 \rangle_{D}
  &=&  {r_{>}} \left[1 + \frac{1}{3} \left(\frac{r_{<}}{r_{>}}\right)^{2}\right]  \label{eq:LD2} \\
  _{D}\langle 2,0 | {|\vec r_{1} - \vec r_{2}|} | 0,2 \rangle_{D} 
  &=& 
  - \frac{1}{5} {r_{>}} \left[ \frac13 \left(\frac{r_{<}}{r_{>}}\right)^{2} - \frac17 \left(\frac{r_{<}}{r_{>}}\right)^{4} \right]  
\end{eqnarray}

In the followings, we list the effective potentials of the Coulomb and linear type potentials 
for higher excited states which we do not include in the calculation:

\begin{eqnarray}
_{S}\langle 2,2 | \frac{1}{|\vec r_{1} - \vec r_{2}|} | 2,2 \rangle_{S} 
  &=& \frac{1}{r_{>}} \left[ 1 + \frac{2}{7} \left(\frac{r_{<}}{r_{>}}\right)^{2} + \frac{2}{7}  \left(\frac{r_{<}}{r_{>}}\right)^{4} \right] \\
  _{S}\langle 2,2 | \frac{1}{|\vec r_{1} - \vec r_{2}|} | 0,0 \rangle_{S}
  &=& \frac{1}{\sqrt 5} \frac{1}{r_{>}} \left( \frac{r_{<}}{r_{>}} \right)^{2}
\end{eqnarray}

\begin{eqnarray}
  _{P}\langle 2,1 | \frac{1}{|\vec r_{1} - \vec r_{2}|} | 2,1 \rangle_{P}  
  &=& \frac{1}{r_{>}} \left[1+\frac{1}{5} \left(\frac{r_{<}}{r_{>}}\right)^{2}   \right] \\
  _{P}\langle 2,1 | \frac{1}{|\vec r_{1} - \vec r_{2}|} | 0,1 \rangle_{P} 
  &=& -\frac{\sqrt 2}{5}  \frac{1}{r_{>}} \left(\frac{r_{<}}{r_{>}}\right)^{2} \\
  _{P}\langle 2,2 | \frac{1}{|\vec r_{1} - \vec r_{2}|} | 2,2 \rangle_{P}
  &=& \frac{1}{r_{>}} \left[ 1 + \frac{1}{7} \left(\frac{r_{<}}{r_{>}}\right)^{2} - \frac{4}{21}  \left(\frac{r_{<}}{r_{>}}\right)^{4} \right] \\
  _{P}\langle 2,2 | \frac{1}{|\vec r_{1} - \vec r_{2}|} | 0,1 \rangle_{P}
  &=&
  %0 \\
  _{P}\langle 2,2 | \frac{1}{|\vec r_{1} - \vec r_{2}|} | 2,1 \rangle_{P}
  =0
 % &=&  0
\end{eqnarray}

\begin{eqnarray}
  _{D}\langle 2,1 | \frac{1}{|\vec r_{1} - \vec r_{2}|} | 2,1 \rangle_{D} 
  &=& \frac{1}{r_{>}} \left[1-\frac{1}{5} \left(\frac{r_{<}}{r_{>}}\right)^{2}   \right] \\
  _{D}\langle 2,1| \frac{1}{|\vec r_{1} - \vec r_{2}|} | 0,2 \rangle_{D}
  &=&  \  _{D}\langle 2,1 | \frac{1}{|\vec r_{1} - \vec r_{2}|} | 2,0 \rangle_{D}
  = 0 \\
  _{D}\langle 2,2 | \frac{1}{|\vec r_{1} - \vec r_{2}|} | 2,2 \rangle_{D}  
   &=& \frac{1}{r_{>}} \left[ 1 - \frac{3}{49} \left(\frac{r_{<}}{r_{>}}\right)^{2} + \frac{4}{49}  \left(\frac{r_{<}}{r_{>}}\right)^{4} \right] \\
  _{D}\langle 2,2 | \frac{1}{|\vec r_{1} - \vec r_{2}|} | 0,2 \rangle_{D} 
  &=& 
  - \sqrt\frac2{35}  \frac{1}{r_{>}} \left(\frac{r_{<}}{r_{>}}\right)^{2} \\
  _{D}\langle 2,2 | \frac{1}{|\vec r_{1} - \vec r_{2}|} | 2,0 \rangle_{D} 
  &=& 
  - \sqrt\frac2{35}  \frac{1}{r_{>}} \left(\frac{r_{<}}{r_{>}}\right)^{2} \\
  _{D}\langle 2,2| \frac{1}{|\vec r_{1} - \vec r_{2}|} | 2,1 \rangle_{D}
  &=& 0
\end{eqnarray}

\begin{eqnarray}
  _{S}\langle 2,2 | {|\vec r_{1} - \vec r_{2}|} | 2,2 \rangle_{S}
  &=& {r_{>}} \left[ 1 + \frac{5}{21} \left(\frac{r_{<}}{r_{>}}\right)^{2} + \frac{2}{77}  \left(\frac{r_{<}}{r_{>}}\right)^{6} \right] \\
  _{S}\langle 2,2 | {|\vec r_{1} - \vec r_{2}|} | 0,0 \rangle_{S}
  &=& - \frac{1}{\sqrt 5} {r_{>}} \left[ \frac13 \left(\frac{r_{<}}{r_{>}}\right)^{2} - \frac17 \left(\frac{r_{<}}{r_{>}}\right)^{4} \right]
\end{eqnarray}

\begin{eqnarray}
  _{P}\langle 2,1 | {|\vec r_{1} - \vec r_{2}|} | 2,1 \rangle_{P}
  &=& {r_{>}} \left[1+\frac{4 }{15} \left(\frac{r_{<}}{r_{>}}\right)^{2}
  +\frac{1}{35} \left(\frac{r_{<}}{r_{>}}\right)^{4}   \right] \\
  _{P}\langle 2,1 | {|\vec r_{1} - \vec r_{2}|} | 0,1 \rangle_{P}
  &=& \frac{\sqrt 2}{5} r_{>} \left[\frac{1}{3} \left(\frac{r_{<}}{r_{>}}\right)^{2}
    - \frac{1}{7}\left(\frac{r_{<}}{r_{>}}\right)^{4} \right] \\  
  _{P}\langle 2,2 | {|\vec r_{1} - \vec r_{2}|} | 2,2 \rangle_{P}
  &=& {r_{>}} \left[ 1 + \frac{2}{7} \left(\frac{r_{<}}{r_{>}}\right)^{2} 
  + \frac{1}{21}\left(\frac{r_{<}}{r_{>}}\right)^{4} -\frac{4}{3\cdot 7 \cdot11} \left(\frac{r_{<}}{r_{>}}\right)^{6} \right] \\
    _{P}\langle 2,2 | {|\vec r_{1} - \vec r_{2}|} | 0,2 \rangle_{P} &=&
    _{P}\langle 2,2 | {|\vec r_{1} - \vec r_{2}|} | 2,0 \rangle_{P} = 0
\end{eqnarray}

\begin{eqnarray}
  _{D}\langle 2,1 | {|\vec r_{1} - \vec r_{2}|} | 2,1 \rangle_{D} 
  &=& {r_{>}} \left[1+\frac{2}{5} \left(\frac{r_{<}}{r_{>}}\right)^{2}
   - \frac{1}{35} \left(\frac{r_{<}}{r_{>}}\right)^{4}   \right] \\
  _{D}\langle 2,1 | {|\vec r_{1} - \vec r_{2}|} | 0,2 \rangle_{D} &=&
  _{D}\langle 2,1 | {|\vec r_{1} - \vec r_{2}|} | 2,0 \rangle_{D} = 0 \\
  _{D}\langle 2,2 | {|\vec r_{1} - \vec r_{2}|} | 2,2 \rangle_{D}
    &=& {r_{>}} \left[ 1 + \frac{4\cdot 13}{3\cdot 7^{2}} \left(\frac{r_{<}}{r_{>}}\right)^{2}
   -\frac{1}{7^{2}}  \left(\frac{r_{<}}{r_{>}}\right)^{4}
   +\frac{4}{7^{2} \cdot 11}  \left(\frac{r_{<}}{r_{>}}\right)^{6} \right] \\
  _{D}\langle 2,2 | {|\vec r_{1} - \vec r_{2}|} | 0,2 \rangle_{D}
  &=& 
   \sqrt\frac2{35}  {r_{>}} \left[\frac13\left(\frac{r_{<}}{r_{>}}\right)^{2}
    - \frac17\left(\frac{r_{<}}{r_{>}}\right)^{4} \right] \\
  _{D}\langle 2,2 | {|\vec r_{1} - \vec r_{2}|} | 2,0 \rangle_{D}
  &=& 
   \sqrt\frac2{35}  {r_{>}} \left[\frac13\left(\frac{r_{<}}{r_{>}}\right)^{2}
    - \frac17\left(\frac{r_{<}}{r_{>}}\right)^{4} \right] \\
  _{D}\langle 2,2 | {|\vec r_{1} - \vec r_{2}|} | 2,1 \rangle_{D} &=& 0 
\end{eqnarray}

\end{document}